\documentclass[aps,PRE,amsmath,amssymb,floatfix]{revtex4-1}
\usepackage{CJK}
\usepackage{graphicx}
\usepackage{dcolumn}
\usepackage{bm}
\usepackage{xcolor}
\usepackage[T1]{fontenc}
\usepackage{lmodern}
\usepackage[colorlinks=true,linkcolor=blue,urlcolor=blue,citecolor=blue]{hyperref}

\begin{document}
\begin{CJK*}{GB}{} 

\title{How to make a giant bubble}

\author{Stephen Frazier }
\author{Xinyi Jiang}
\author{Justin C. Burton}
\email{justin.c.burton@emory.edu}

\affiliation{Department of Physics, Emory University, Atlanta, Georgia 30322, USA}

\date{\today}

\begin{abstract}
Using mixtures of soap, water, and long chain polymers, free-floating soap bubbles can be formed with volumes approaching 100 m$^3$. Here we investigate how such thin films are created and maintained over time. We show how the extensional rheology is the most important factor in creating the bubble, and how polydispersity in molecular weight of the solvated polymers leads to better performance at lower concentrations. Additionally, using IR absorption, we measure soap film thickness profiles and film lifetimes. Although the initial thickness mostly depends on the choice of detergent, polymers can dramatically increase film lifetime at high molecular weights and high concentrations, although such high concentrations can inhibit the initial film formation. Thus, the ideal concentration of polymer additives for making giant bubbles requires a robust viscoelastic rheology during extension, and is aided by long film lifetimes during gravitational drainage and evaporation.
\end{abstract}

\maketitle
\end{CJK*}

\section{Introduction}

For the last 2 decades, giant bubble enthusiasts have been creating soap film bubbles with ever increasing volumes. As of 2019, the world record for a free-floating soap bubble stands at 96.27 m$^3$ \cite{record2015}. For a spherical bubble, this corresponds to a diameter of 5.7 m and a surface area of 101 m$^2$. A simple glance at the multitude of colors from the reflecting bubble film suggests a film thickness of order a few microns. Thus, the extension-to-thickness ratio of this film is nearly $5\times10^5$, which is quite staggering considering that a single hole can lead to the film's demise. How are such large films created, and how do they remain stable? These questions pose interesting fluid mechanics problems, but are also important considering the vast role that soap bubble films have and continue to play in physics research \cite{Goldstein2010,Rivera1998,Ristroph2008,Salkin2016,Poulain2018} and education \cite{Isenberg1981,Ramme1992,Ramme1997}. Additionally, the formation, mechanics, and stability of foams with various additives is an important environmental problem. Such foams contribute to long-lasting pollution in rivers and waterways contaminated with industrial run-off \cite{Schilling2011,toxicfoam}. 

For those interested in making giant bubbles, the Soap Bubble Wiki \cite{soapwiki} contains a wealth of empirical information and recipes for optimal bubble solutions. Most solutions use industrial dish detergents as a surfactant for the soap films. For soap films supported solely by surface tension forces, one may estimate the maximum size of a bubble based on the Bond number, Bo $\equiv\rho g z R_b/\sigma$, where $R_b$ is the radius of the bubble, $z$ is the thickness of the film, $\rho$ is the liquid density, $\sigma$ is the surface tension, and $g$ is the acceleration due to gravity \cite{Lautrup2011}. The Bond number measures the ratio of gravitational forces to surface tension forces. Assuming Bo = 1 for films with $d\approx$ 2 $\mu$m, the maximum size is $R_b\approx1.5$ m. Although Marangoni forces can help support larger films, up to a few meters \cite{deGennes2001,deGennes2004,Cohen2017}, additional forces are necessary to make world-record giant bubbles.

The key ingredient is the addition of long chain polymer molecules. Two of the most common types are guar gum and polyethylene oxide (PEO). When solvated in water, guar forms long polysaccharide chains of polydisperse molecular weight, typically between 0.25-5.0 $\times$ 10$^6$ g/mole \cite{Mudgil2014}. PEO is more well-controlled, as samples of monodisperse molecular weights can be obtained, although most industrial samples are polydisperse. Both additives drastically alter the viscoelastic rheology of the soap solution in different ways, and are partially confined in soap films since their radius of gyration is fractions of a micron. Despite this wealth of information, the physics of making giant bubbles with polymer solutions is poorly understood.  Properties such as molecular weight, concentration, and polydispersity are exceedingly important, as well as the solution's overall rheology and choice of surfactant. It is no wonder a complete on-line Wiki exists that is devoted to the fine-tuning of giant bubble recipes. 

In this paper we identify some of the underlying physical mechanisms that give rise to giant bubbles. We study both the shear and extensional rheology of common solutions of water, surfactant, and PEO or guar. The most robust solutions for making bubbles have intermediate concentrations and a polydisperse mixture of polymers of various molecular weights, allowing a large volume of liquid to be continuously drawn into a film without breaking. Additionally, we measure the thickness of soap films over time using IR absorption and find that the lifetimes increase at larger concentrations than those often used in bubble solutions, indicating that polymers may also enhance film longevity after evaporation and gravitational drainage have occurred. We suggest that an optimal polymer solution making giant bubbles comes from a combination of the extensional rheology of isolated chains, and cooperative interactions between polymers of differing molecular weights.

\section{Experimental Methods}

The basic solutions used in our experiments consisted of a mixture of detergent and water. Unless otherwise noted, all solutions were a mixture of 4\% v/v Dawn Pro Dish Detergent and deionized water. For some experiments, we used a detergent common in many soap film experiments, sodium dodecyl sulfate (SDS), which was obtained from MilliporeSigma. Long-chain polymers were then added to each soap solution. We measured the surface tension of all soap solutions using axisymmetric drop shape analysis \cite{Burton2010}. Regardless of polymer concentration, all solutions had a surface tension of $32\pm2$ mN/m. Guar powder was obtained from MilliporeSigma. The powder was directly added to the soap solution, then magnetically stirred at 50$^\circ$C for 4 hrs to allow for full dissolution of the particles. Guar is often dissolved into an alcohol slurry first to prevent clumping. However we found that heating and stirring worked just as well. Solutions were ready to use after cooling to room temperature (22$^\circ$C). For making giant bubbles, guar is often added in concentrations ranging from 1.0-2.0 g/l \cite{soapwiki}, although we explored a wider range in our experiment.

PEO was also added to soap solutions. First, we used a common industrial PEO lubricant for making giant bubbles, J-Lube (Jorgensen Labs), which is composed of 25\% polydisperse PEO (up to 8 $\times$ 10$^6$ g/mole) and 75\% sucrose. We did not characterize the composition of the J-Lube sample beyond information from the supplier. For making giant bubbles, J-Lube is often added in concentrations of 0.1-0.4 g/l \cite{soapwiki}, 25\% of which is actually PEO, so the concentrations are much lower than guar. In addition to commercial PEO, we used monodisperse samples of PEO from MilliporeSigma with molecular weights of 0.1M, 0.6M, 1.0M, 2.0M, 4.0M, and 8.0M g/mole, where M = 10$^6$. These are viscosity-averaged molecular weights ($M_v$), which lies between the number-averaged and weight-averaged molecular weights.  The polydispersity index was not available for these samples. All powdered samples of PEO and guar were stored in opaque containers and placed in the refrigerator until use. Some PEO was intentionally degraded by aging for 6 months in room temperature conditions, as noted. 

\begin{figure}
\includegraphics[width=4.5 in]{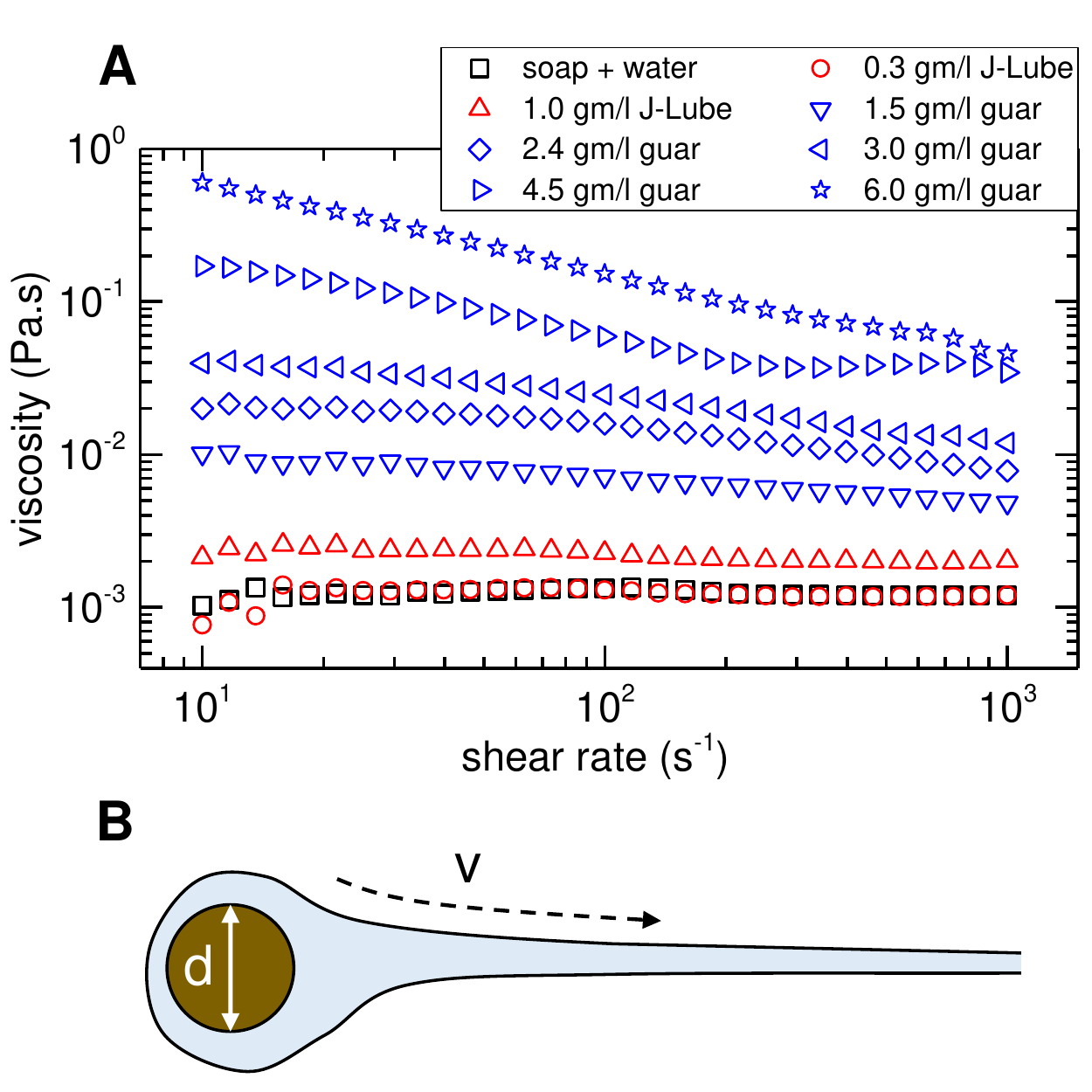}
\caption[]{(\textbf{A}) Shear rheology of the soap + water + polymer solutions at 25$^{\circ}$. The black squares are the control (no polymer), and the blue and red points represent soap solutions with polymers added in the indicated concentrations. (\textbf{B}) Schematic of a soap film being pulled at characteristic velocity $v$ from a rope with a circular cross-section of diameter $d$. Away from the rope, the flow is mostly extensional.
\label{rheology}} 
\end{figure}

\section{Shear rheology and extensional properties}

For an initial characterization of the bubble solutions, we employed steady-state rheometry to measure the concentration and shear rate dependence of the shear viscosity. We used a TA Instruments AR2000 with a parallel-plate configuration and a 0.5 mm gap. A Peltier plate was used to control the temperature of the sample. Figure \ref{rheology}A shows the results from various concentrations of guar and J-Lube, as well as the pure soap solution. For the J-Lube solutions, the viscosity varies by less than a factor of 2, even for concentrations larger than those used for making giant bubbles ($\approx$ 0.1-0.4 g/l). All the guar solutions tested showed some degree of shear-thinning, and the shear viscosity increased dramatically with concentration at low strain rates. Given that both 1.5 g/l and 2.4 g/l are concentrations used for giant bubbles, and their viscosities are approximately 10 times that of the J-Lube solutions, we conclude that the shear viscosity is not the leading factor in determining a solution's ability to create giant bubbles. 

Figure \ref{rheology}B shows a representation of a soap film being pulled from a rope of diameter $d$ with a characteristic velocity $v$. Near the rope where the velocity is zero at the boundary, the fluid is sheared with a characteristic shear rate $\dot{\gamma}\approx v/d$. For a 1 mm rope with a film velocity of $U$ = 1 m/s, the shear rate is 1000 s$^{-1}$. This is an upper bound and the highest shear rate we tested in our shear rheology. However, away from the rope, the flow is mostly extensional due to slow variation in film thickness in the pulling direction \cite{Nierop2008}. Thus, we expect the bulk extensional viscosity to be more important for creating giant bubbles. For solutions of long chain polymer molecules, it is well-known that the extensional properties change dramatically at very low concentrations where changes in the shear viscosity are nearly immeasurable \cite{Zimm1956,Wagner2005,Clasen2006,Palangetic2014,Giudice2017}. 

To characterize the extensional viscosity of each sample, we used high-speed video to analyze the falling of droplets under gravity, and the dynamics of the viscoelastic thread that they pulled in their wake. Although previous studies have used the break-up rate of a liquid bridge to characterize non-Newtonian extensional rheology \cite{McKinley2002,Dinic2017,Dinic2019}, we chose to measure the length of the viscoelastic thread prior to its final rupture since this is more analogous to the continuous pulling of a soap film. Bulk liquid was filled into a glass burette, and the flow rate was adjusted so that one drop fell every 10 s. To image the drops, we used a Vision Research Phantom v7.11 camera at 2000 frames-per-second with a Tonkia macro lens, resulting in a resolution of 70 pixels/mm. The liquid drops were illuminated from behind with a 150 W tungsten-halogen lamp. 

\begin{figure}
\includegraphics[width=5 in]{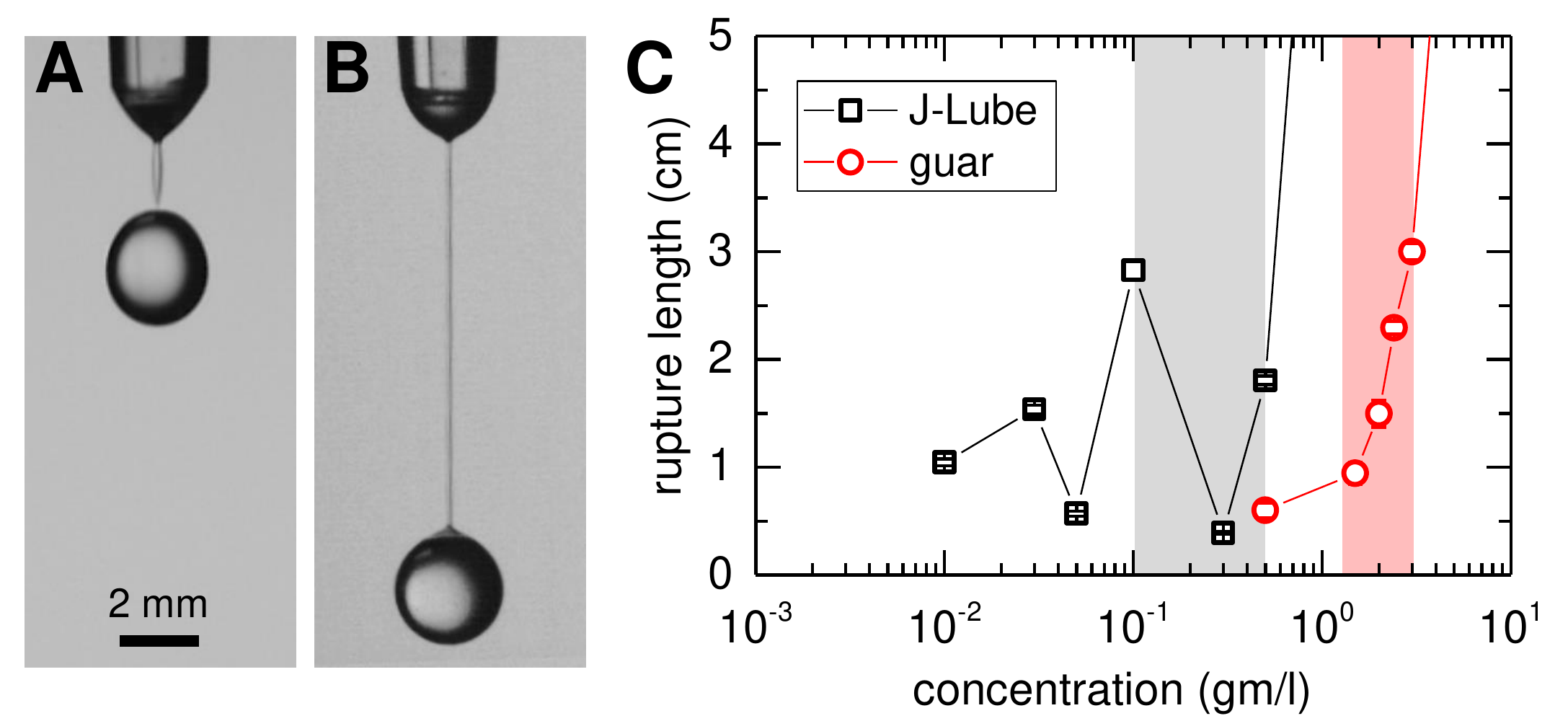}
\caption[]{(\textbf{A}) Image of a soap + water solution at the moment the connecting thread ruptures near the top of the drop. (\textbf{B}) A solution with 3.0 g/l of guar added. The thread is highly extended, and will reach $\approx$ 3 cm in length before rupture. The scale bar applies to both images. (\textbf{C}) Thread rupture length vs. polymer concentration for J-Lube and guar. The gray and light red rectangles indicate the range often used to make giant bubble solutions. Error bars represent the standard deviation of multiple drops, although the experiments were extremely repeatable.
\label{guar_jlube}} 
\end{figure}

Figure \ref{guar_jlube}A-B shows images of two drops immediately after detachment from the joining thread. Without the addition of polymers, the thread resembles that of pure water and other low-viscosity pure liquids \cite{Burton2007}. When 3.0 g/l of guar is added, the thread length increases dramatically. At higher concentrations, the length of the thread prior to rupture seems to diverge, and the drop leaves the imaging region before the thread ruptures. While this behavior is monotonic for the guar solutions, the PEO-based J-Lube solutions oscillate between long and short threads before finally diverging in length around 0.7 g/l (Fig.\ \ref{guar_jlube}B). We attribute this non-monotonic behavior to the formation of ``beads-on-a-string'' during extension in the lower-viscosity J-Lube solutions \cite{Bhat2010,Wagner2005}. Guar solutions typically have a 10 fold larger shear viscosity, which suppresses the formation of beads through inertial forces \cite{Bhat2010}. Rupture of the thread can then occur where beads are connected to the thread rather than at the main drop. 

Although the particular placement of the oscillations shown in Fig.\ \ref{guar_jlube}C are likely specific to our dripping experiment, the data shows that the addition of either J-Lube or guar has a significant effect on the extensional properties of the solutions for concentrations used to make giant bubbles. However, the molecular weight distribution of both polymer sources is unknown. Both contain high molecular weight ($> 1M$) components, but this alone is not sufficient, and is potentially detrimental to making giant bubbles. Many enthusiasts prefer to ``age'' their PEO in the powder form for many weeks or months prior to usage \cite{soapwiki}. Solutions of newly-purchased J-Lube and other industrial sources of PEO are often very sensitive to concentration and can become quite ``stringy'' and elastic. This is evident in the Fig.\ \ref{guar_jlube}B where the rupture length begins to diverge. 

Aging the PEO in solid-state involves photo-induced or thermal degradation where larger chains break into smaller ones, and the distribution of molecular weight broadens \cite{Claire2009,Morlat2001}. Degradation can also occur in solution as a result of aging or strong fluid flow \cite{Dupas2012,Buchholz2004,Muller1992}. The concentration at which chains begin to overlap and interact, denoted by $c^*$, depends on the molecular weight. At lower concentrations, the rheology of very dilute solutions depends strongly on the nonlinear extensional properties of individual, solvated polymer chains. Thus we investigated the extensional properties of soap solutions with monodisperse PEO of different molecular weights. 

\begin{figure}
\includegraphics[width=6.3 in]{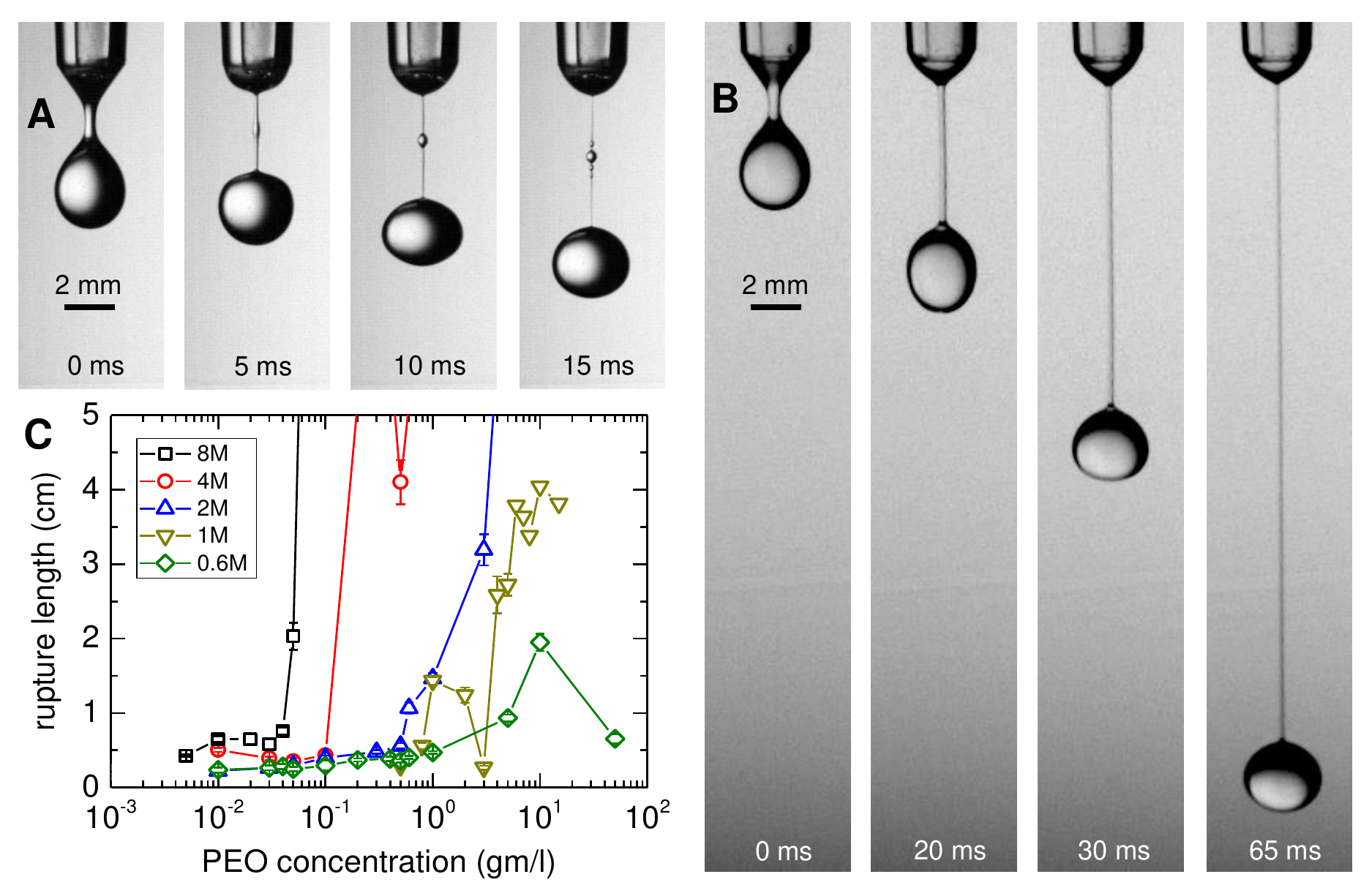}
\caption[]{(\textbf{A}) Images of a falling drop with 50.0 g/l of 0.6M PEO. The formation of a distinct bead induces rupture of the thread. (\textbf{B}) Images of a falling drop with 0.05 g/l of 8.0M PEO. The drop exits the viewing area before rupture of the thread. (\textbf{C}) Rupture length vs. concentration for 5 different molecular weights of PEO. Error bars represent the standard deviation of multiple drops. 
\label{drop_fig}} 
\end{figure}

\begin{table}[]
\caption{Molecular weights of PEO used in the experiments and their corresponding radii of gyration ($R_g$) and overlap concentration ($c^*$) assuming they are well-solvated.  The quantities $R_g$ and $c^*$ were computed as described in the text and references \cite{Holyst2009,Devanand1991,Ying1987}.}
\centering
\begin{tabular*}{3.0 in}{l @{\extracolsep{\fill}} cr}
     \hline
     PEO MW                &$R_g$ (nm)  &$c^*$ (g/l)     \\ 
           
     \hline
     0.1M               &16       &9.90	              	 \\
    0.6M	    &45         &2.63     	     	    \\
     1.0M	          &60       &1.80	       	    \\
     2.0M          &90       &1.08	       	    \\
     4.0M           &135	     &0.65	           \\
     8.0M       &202       &0.39	       	    \\
     \hline
     \label{tab1}
\end{tabular*}
\end{table}

Table \ref{tab1} lists the molecular weight, radius of gyration ($R_g$), and overlap concentration ($c^*$) for the samples used in our experiments. The radius of gyration for PEO was calculated using the formula $R_g=0.02 M_v^{0.58}$ [nm], in accordance with references \cite{Holyst2009,Devanand1991,Ying1987}, and the overlap concentration was calculated as $c^*=M_v/(4\pi R_g^3 N_A/3)$ [g/m$^3$], where $N_A$ is Avogadro's number. We have assumed that the polydispersity of each sample is small enough so that the viscosity and weight-averaged molecular weights are the same (i.e. $M_v=M_w$).

In analogy to Fig.\ \ref{guar_jlube}, Fig.\ \ref{drop_fig}A-B shows images of falling drops of PEO solutions of different molecular weights. For lower molecular weights, 1.0M or less, the increase in the rupture length coincides more closely with $c^*$, indicating that polymer blob overlap and transient entanglements may contribute to the extensional rheology. High molecular weights, above 2.0M, lead to divergent rupture lengths above a certain concentration less than $c^*$ (Tab.\ \ref{tab1}). The nonlinear rheology is apparent; an 8.0M solution produces extremely long threads at concentrations 1000 times less than a 0.6M solution. This shows that for very high molecular weights, the entropic cost of extension for individual chains in the flow leads to viscoelastic properties and the formation of long threads. 

One major difference between most of the data shown in Fig.\ \ref{drop_fig} and the data for J-Lube shown in Fig.\ \ref{guar_jlube} is the nearly monotonic behavior of the monodisperse PEO, and the oscillations in the polydisperse J-Lube solutions. A polydisperse solution will certainly contain a broader range of relaxation times in its rheological response \cite{Cross1969,Kulicke1983,Bhattacharjee2002,Ye2003,Ye2005,Palangetic2014}, possibly giving rise to a non-monotonic extensional response. The rheology of a solution at very low concentrations would depend mostly on the longest polymers in the polydisperse mixture. However, one would expect that as a collection of polymers ages or degrades, the individual polymers would not lengthen, and the distribution would only widen toward shorter lengths.

\begin{figure}
\includegraphics[width=6 in]{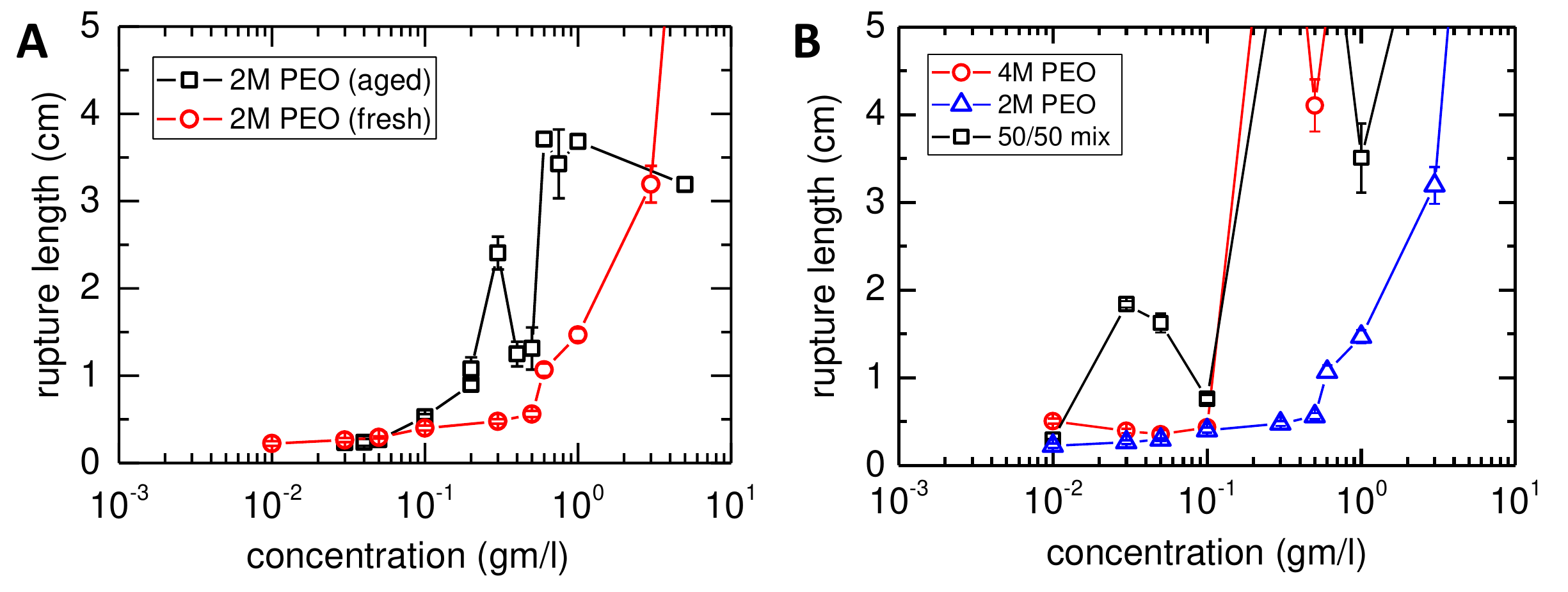}
\caption[]{(\textbf{A}) Rupture length vs. concentration for freshly-purchased 2M PEO and for 2M PEO that has been aged for 6 months at room temperature. (\textbf{B}) Rupture length vs. total concentration for 4M PEO, 2M PEO, and a 50/50 mixture of each sample. Error bars represent the standard deviation of multiple drops in both graphs. 
\label{poly_mix}} 
\end{figure}

Our results with aged and polydisperse mixtures suggest a more cooperative behavior. Figure \ref{poly_mix}A shows the rupture length for falling drops in a freshly-purchased sample of 2M PEO, and a sample that has been aged for 6 months at room temperature in an opaque bottle. Surprisingly, the aged sample shows a stronger extensional response at lower concentrations. We can also observe the same effect by simply mixing a 2M and 4M PEO sample together. For the same total weight of polymer, the mixture displays an increase in the length of the thread at lower concentrations than either monodisperse sample. This behavior is only possible if there are interactions between the longer and shorter chain polymers, i.e. they can not be considered as dilute, individual molecules. 

Although the exact nature of these interactions is unclear, we hypothesize that some degree of clustering of the longer-chain polymers may give rise to the measured results. The smaller polymers may act as depletants, leading to a weak, attractive force between the longest and largest polymers in the solution. Transient entanglements between long polymers in close proximity could then lead to extensional resistance at low bulk concentrations, with an ultimate elastic resistance limited by the strength of the transient entanglement. To our knowledge, evidence for this behavior has not been reported in the literature, and is left for future studies focused on extensional rheology. Nevertheless, it is evident that the polydispersity in molecular weight makes the extensional properties of bubble solutions less sensitive to changes in concentration, and thus more robust for making giant bubbles. 

\section{Lifetime and thickness of soap films}

There is approximately 300 ml of liquid in a giant spherical bubble with a surface area of 100 $m^2$ and an average thickness of 3 $\mu$m. Having the right balance of extensional properties that resist thinning and breaking but allow for sufficient flow rates is one part of the puzzle in making giant bubbles. These bubbles must also survive for a considerable amount of time. However, the thinning of soap films is quite complex \cite{Mysels1959}. Although vertical soap films will drain and thin under the influence of gravity, evaporation also plays a key role \cite{Champougny2018}. Low humidity and high temperatures are well-known barriers to long-lived, giant bubbles \cite{soapwiki}. Assuming a constant thickness, both the volume of the liquid and mass evaporation rate should be proportional to the surface area of the film, so to first approximation, evaporation should affect both small and large films in the same manner, provided that the diffusion of water vapor away from the surface is similar. 

Marangoni forces and film stretching can also significantly affect the lifetime of soap films without polymers \cite{Saulnier2014}. The type of surfactant and its overall concentration will determine the flow profile in the film by altering the the degree of ``slip'' at the liquid-air interface \cite{Naire2000,Berg2005,Schwartz1999}, and positive vertical gradients in the surfactant concentration at the surface will pull upwards on the film \cite{deGennes2001,deGennes2004}. Furthermore, the 2D surface rheology can influence the flow and overall drainage of the film \cite{Langevin2000,Seiwert2014,Sonin1994,Nierop2008}.  Finally, polymers have been shown to affect the initial thickness of the film at high concentrations, near or above $c^*$, displaying strong deviations from Frankel's law \cite{Addad1991,Adelizzi2004,Bruinsma1992,Adelizzi2004}.  Polymer-surfactant interactions may also be important \cite{Adelizzi2004,Addad1994}, these systems require more quantitative studies. Nevertheless, nearly all of these experiments use soap films of a few centimeters, well below the size of giant bubbles.

\begin{figure}
\begin{center}
\includegraphics[width=5 in]{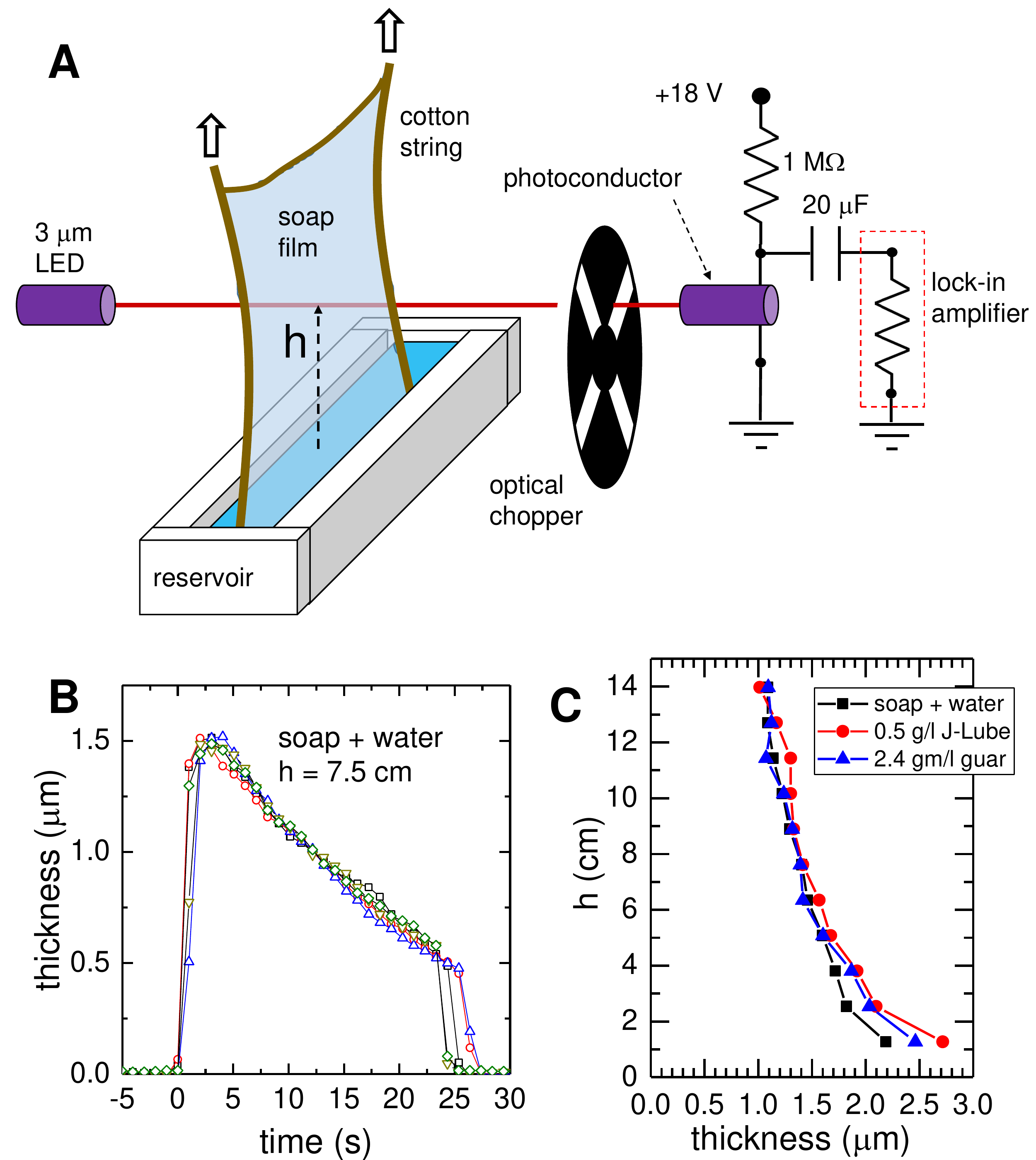}
\caption[]{(\textbf{A}) Schematic of the experimental apparatus used for thickness measurements. The infrared light from the LED is modulated by the optical chopper and detected using a lock-in amplifier. The height of the measurement ($h$) can be adjusted for each experiment. (\textbf{B}) Three separate experiments of the thickness vs. time at $h$ = 7.5 cm for a soap + water film with no polymers. The films popped after $\approx$ 25-30 s. (\textbf{B}) Film thickness profiles at $t$ = 5 s for three different bubble solutions. Each data point is a separate experiment. The thickness error for each measurement is $\approx$ 100 nm, and is mostly due to the variation between experiments. 
\label{film_cartoon}} 
\end{center} 
\end{figure}

It is not clear if large soap films with polymers last longer by producing thicker films, slowing gravitational drainage, or slowing evaporation. To answer some of these questions, we performed time-dependent thickness measurements of soap films using a custom-built infrared absorption apparatus. Our design is similar to those used in previous measurements of soap films \cite{Wu2001}. Figure \ref{film_cartoon}A shows a schematic of the setup. Two cotton strings (1 mm diameter) were anchored in a liquid reservoir containing the soap solution, connected by a third cotton string at the top. The strings were submerged into the solution and then quickly raised and made tight in order to produce a rectangular soap film with dimensions 10 cm $\times$ 15 cm. The film was allowed to drain under gravity until it popped. 

An infrared LED (Boston Electronics) with a peak wavelength $\lambda$ = 3 $\mu$m was used as a semi-collimated source. The light was passed through the center line of the film and detected by a photoconductor (Thor Labs) on the other side. The signal was modulated at 600 Hz by an optical chopper and detected by a lock-in amplifier (SRS 830, 1 M$\Omega$ input impedance). The impedance of the photoconductor is approximately 1 M$\Omega$, but varies by a small amount based on the intensity of the incoming light. The capacitor shown in the figure is used to block the DC component of the signal. The optical elements were mounted on an adjustable stage so that the height of the beam from the liquid bath, $h$, could be varied for each experiment. A large acrylic box was placed around the entire experimental apparatus in order to control for ambient air currents and to keep the humidity stable. 

Water has a strong, sharp absorption peak at $\lambda$ = 3 $\mu$m, so that the exponential extinction length is $z_0\simeq 0.9$ $\mu$m \cite{Wu2001}. Taking into account the reflections at the incident interface and within the film (index $n$ = 1.17), the ratio of the transmitted to incident intensity is
\begin{eqnarray}
&\dfrac{I_T}{I_0}=\dfrac{(1-R)^2 e^{-z/zo}}{1-R^2e^{-2z/zo}},
\end{eqnarray}
where $R=(n-1)^2/(n+1)^2$ and $z$ is the local film thickness. However, as discussed in reference \cite{Wu2001}, $R$ can be safely ignored since $R=0.0061$ for $n=1.17$. Thus, after verifying the linearity of our optical detector using optical density filters to reduce the intensity of the incident beam, we used the following relationship to convert the voltage measured to the film thickness:
\begin{eqnarray}
&V(z)=V_0 e^{-z/zo},
\end{eqnarray}
where $V_0$ is the voltage measured in the absence of the film.

\begin{figure}
\begin{center}
\includegraphics[width=6 in]{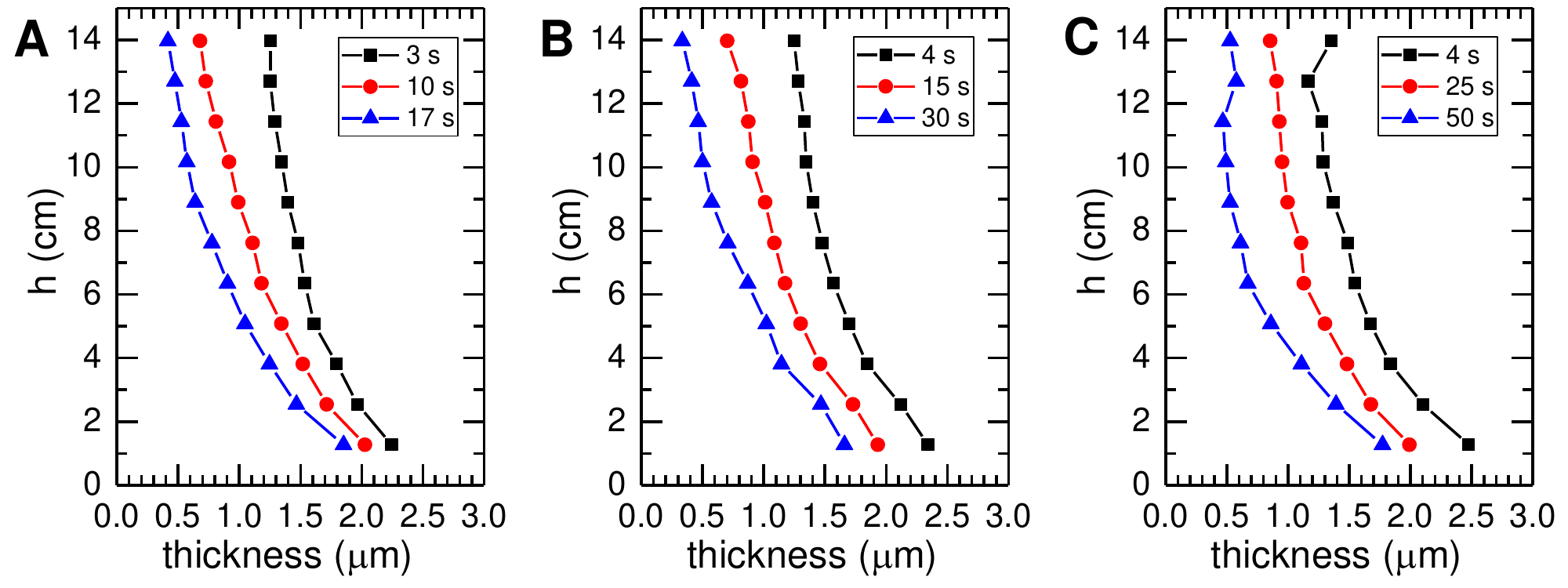}
\caption[]{Film thickness profiles at 3 different times for a soap + water solution (\textbf{A}), a solution with 0.5 g/l of 1M PEO (\textbf{B}), and a solution of 0.5 g/l of 8M PEO (\textbf{C}). Each data point is a separate experiment. The thickness error for each measurement is $\approx$ 100 nm, and is mostly due to the variation between experiments.  
\label{profiles}} 
\end{center} 
\end{figure}

Figure \ref{film_cartoon}B shows thickness vs. time data for three separate soap films with no polymers at $h$ = 7.5 cm. The initial rise is the raising of the cotton strings to create the film, and the slow decay is the thinning of the film until it eventually pops at $t\approx$ 25 s. We found that there was often more variability in the final film lifetime than in the initial thickness and drainage rate. This may be expected since the film can pop due to an instability that forms anywhere in the film. Most films popped when some portion became thinner than 500 nm, although some films with polymers reached smaller thicknesses. Perhaps surprisingly, the overall film thickness did not change much with the addition of polymers. Figure \ref{film_cartoon}C shows initial profiles of the film thickness vs. height for pure soap + water, a solution with J-Lube, and a solution with guar. Data for each height was taken with different, identical soap films since the apparatus had to be raised and lowered between each experiment. As will be shown later, the film thickness depended much more strongly on the choice of surfactant rather than the addition of any polymers.

\begin{figure}
\begin{center}
\includegraphics[width=6.0 in]{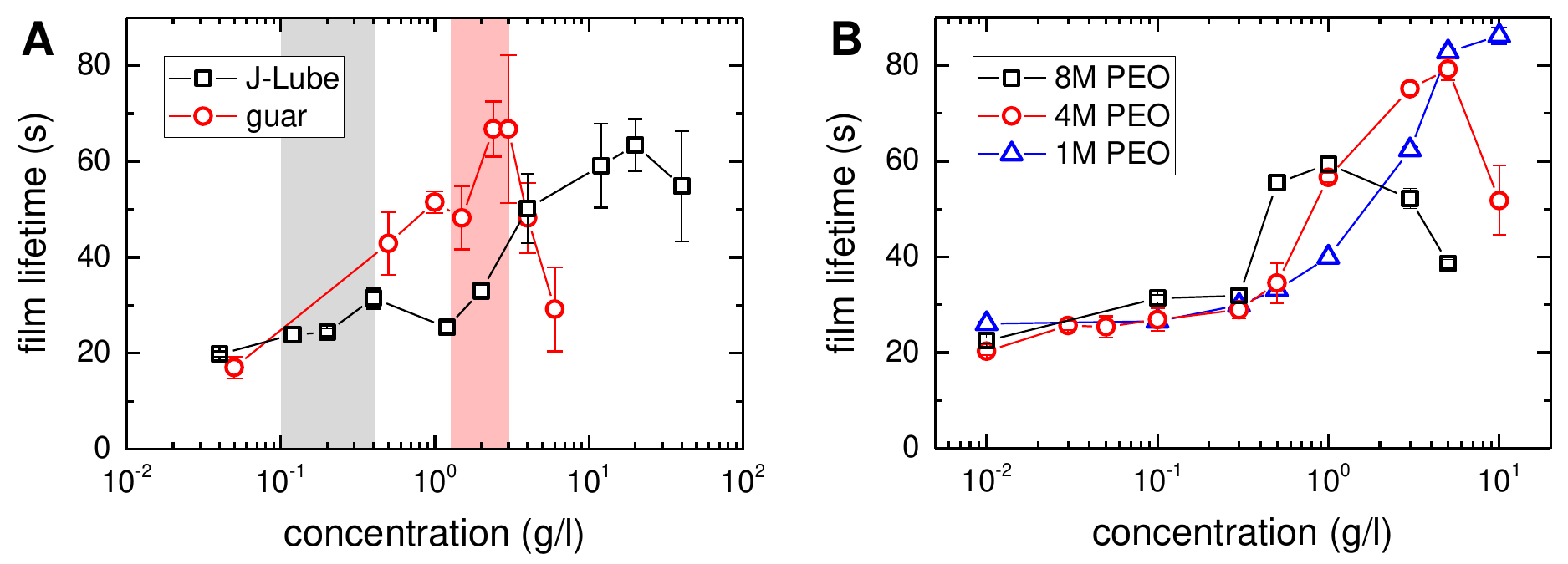}
\caption[]{(\textbf{A}) Film lifetime vs. concentration for both J-Lube and guar bubble solutions. The gray and light red boxes indicate the concentrations typically used to make giant bubbles. (\textbf{B}) Film lifetime vs. concentration for monodisperse PEO polymer solutions. Reported error bars represent the variation among repeat experiments.
\label{life_vs_conc}} 
\end{center} 
\end{figure}

The overall lifetime of the film strongly depended on the presence of polymers in the solution. Figure \ref{profiles} shows the thickness profiles at 3 different times for soap + water (A), and 0.5 g/l solutions of 1M and 8M PEO. The qualitative shape of the profiles was basically the same, although there was some variation near the top of the film for the 8M PEO sample (Fig.\ \ref{profiles}D), possibly due to a two-dimensional ``beads-on-a-string'' effect where thick and thin regions exist simultaneously.  However, the concentrations required to significantly increase the lifetime were distinctively larger than that needed to produce long, viscoelastic threads in falling drops, as shown in Fig.\ \ref{drop_fig}. The slow, linear drainage and long timescale (more than 10 seconds) suggests that the flow profile in the film is more parabolic than uniform, possibly due to an immobile surfactant boundary for large concentrations of soap \cite{Mysels1959,Schwartz1999,Berg2005,Naire2000}. In this case, the polymers experience a simple Poiseuille-type flow since shear-thinning behaviors require higher concentrations (Fig.\ \ref{rheology}) where individual polymer chains overlap above $c^*$. If the velocity profile in the film was more like ``plug'' flow, then we would expect and exponential dependence on time \cite{Debregeas1998}.

This concentration and molecular weight dependence can be seen in Fig.\ \ref{life_vs_conc}. The optimal concentrations for making giant bubbles (J-Lube and guar; gray and light red regions in Fig.\ \ref{life_vs_conc}A) show that film lifetime is not the most important property since the lifetime for guar is much longer than for J-Lube at the optimal concentrations. However, the correlation between the extensional properties and the maximum lifetime for guar solutions may be what makes it an excellent choice for making large, stable soap bubbles. The lifetime also depends on the molecular weight of the individual polymers. Figure \ref{life_vs_conc}B shows that the lifetime of a film typically increases with concentration, then decreases at high concentrations. For 1M PEO, the lifetime can be longer than 80 s, whereas a film with 8M PEO never lives longer than 60 s. We suggest that the increase in the shear viscosity of the solution alongside the increase in extensional properties for lower molecular weight polymers is the primary reason for such long lifetimes, whereas for high molecular weight polymers, the film is more elastic, and may be more susceptible to breakage and failure.

\begin{figure}
\begin{center}
\includegraphics[width=7 in]{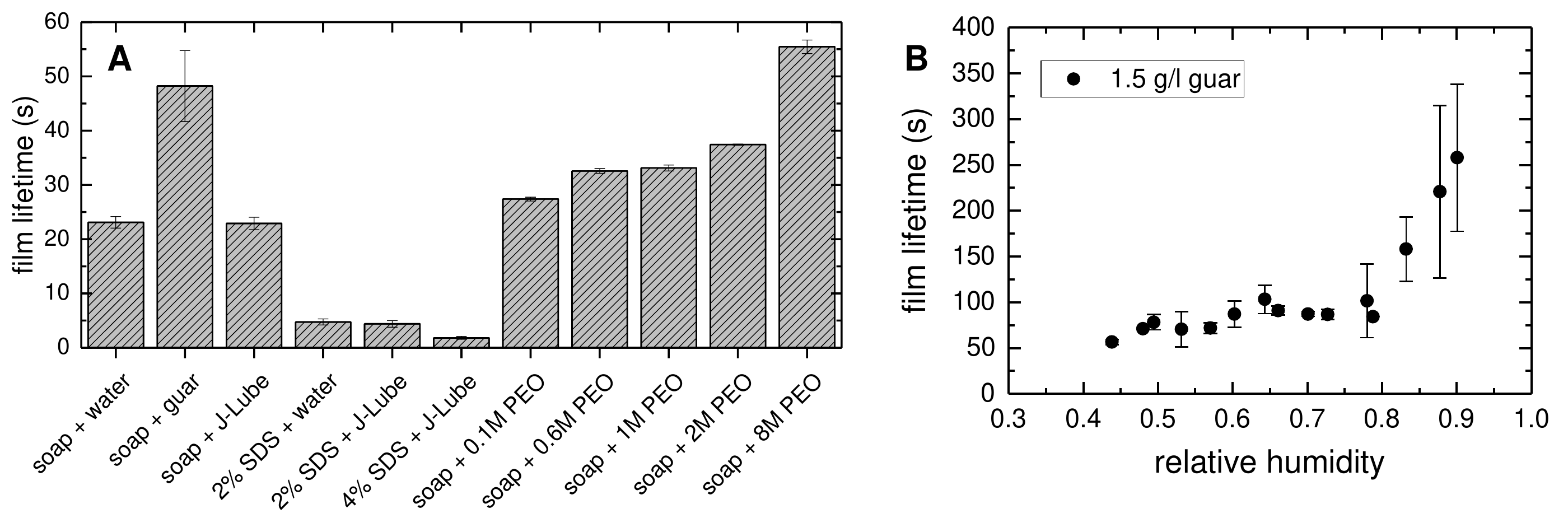}
\caption[]{ (\textbf{A}) Average lifetimes of films formed from different bubble solutions. The guar concentration used was 2.4 g/l, and was 0.5 g/l for all J-Lube and PEO solutions. Note that J-Lube is only 25\% PEO by weight. (\textbf{B}) Film lifetime vs. relative humidity for a 1.5 g/l guar solution. Error bars for all data represent the variation among multiple experiments. 
\label{lifetime_bargraph}} 
\end{center} 
\end{figure}

A summary of lifetimes from various bubble solutions is shown in Fig.\ \ref{lifetime_bargraph}A. Also shown are lifetimes for films made with a commonly used detergent for making soap films, SDS. For small SDS soap films of order $\sim$ 1-2 cm$^2$, lifetimes of the films can reach 30 s or more \cite{Schwartz1999,Bhamla2017}, often with the help of Marangoni stresses. We found that for films of order $\sim$ 100 cm$^2$, SDS resulted in extremely short lifetimes. Even though the volume concentrations are similar, Dawn Pro dish detergent dramatically increases the stability of the film. The main conclusion from these data is that although lifetime itself does not determine a bubble solution's ability to make giant bubbles, it is certainly not a detriment. As soon as a film is pulled, gravity will act to drain the liquid from parts of the film. Additionally, the water will begin to evaporate from the huge surface area, also thinning the film. The latter will increase the concentration of any polymer, and help to increase the lifetime and slow the drainage. As Fig.\ \ref{life_vs_conc}B shows, a polydisperse mixture of molecular weights can help to increase the lifetime over a broader range of concentrations. 

Finally, as is known to many bubble enthusiasts, evaporation will decrease the lifetime of bubbles no matter which polymer is used. A recent experimental study showed that the time to rupture after the formation of a centimeter-scale soap film strongly depends on humidity \cite{Champougny2018}. In our experiments the effect is dramatic. This  is shown in Fig.\ \ref{lifetime_bargraph}B for a 2.4 g/l guar solution. The relative humidity (RH) in the experimental enclosure was increased using a cup of warm water exposed to the air. A slow increase in the lifetime was observed for low humidity, and then at $\approx$ 75\% RH, the lifetime began to increase, reaching almost 250 s at 90\% RH. Considering this significant increase, we can conclude that much of the film thinning occurs through evaporation. As mentioned previously, the liquid volume and mass evaporation rate both scale with surface area, so for larger soap films, the removal of the evaporative boundary layer through convection or ambient wind is likely to be important. It is no wonder that many bubble enthusiasts prefer cool, humid days for making giant bubbles. 

\section{Conclusion}

In conclusion, we have investigated the properties of surfactant and polymer solutions commonly used to make giant bubbles with surface areas approaching 100 m$^2$ and thicknesses of a few microns. The most important additives for making these bubbles are hydrophillic, high molecular weight polymers, such as guar gum and PEO. We found that it is the extensional rheology of these dilute, polymer solutions that is the most important factor in creating the films. Although PEO is mildly surface-active and can absorb to the air-water interface \cite{Cao1994,Adelizzi2004}, we are not aware of any studies on its surface rheology. Given that the range of polymer concentrations used for giant bubbles agree well with the bulk extensional properties, we suspect that surface viscoelastic effects play a minor role in the formation of giant bubbles.

Assuming a pulling velocity of $U\approx$ 1 m/s, the capillary number associated with pulling a giant bubble is Ca $\equiv U\eta/\sigma\sim$ 0.05, were $\eta$ is the bulk viscosity, assuming it is Newtonian. This is much larger than those used in typical lab experiments \cite{Mysels1959,Adelizzi2004,Berg2005}. If a soap film is pulled too fast, then surfactants do not have time to migrate to the newly-created free surface and stabilize the film \cite{deGennes2001}. Additional elastic forces from added polymers are thus necessary to prevent film breakage during the formation of a giant bubble. Although the lifetime of the film does increase with polymer concentration, this is a secondary effect that occurs at higher concentrations, and may influence the stability of the film during evaporation and thinning. Other than the concentration, molecular weight, and polydispersity of the polymers, there are other factors that are often tuned to make the best bubble solutions, such as pH buffering, which is usually done with the addition of baking powder. Although, we expect this mostly affects the agglomeration of solvated polymers, and the ensuing shelf life of the solution.

One main finding of this work that requires further investigation is the cooperative influence of polydisperse solutions. Mixtures of two different molecular weights, or aged solid samples of polymers would produce noticeable extensional properties at concentrations much lower than required in monodisperse solutions. This can not be explained by the properties of dilute, polymer chains. In polydisperse polymer melts, cooperative effects have been identified, such as nematic phase separation \cite{Zhang2018}. Interactions between the surfactant micelles and polydisperse polymers may also lead to a rich set of emergent interactions \cite{Clegg1994}, which may also depend on the ambient flow of the solvent \cite{Helfand1989,Milner1991}. Dedicated extensional rheology experiments and molecular dynamics simulations may help to shed light on the nature of these cooperative effects in future studies. 

\begin{acknowledgments} 
This work was supported by the NSF DMR Grant No. 1455086. 
\end{acknowledgments}

\end{document}